# A Thulium-Silicon Hybrid Microdisk Laser

Khadijeh Miarabbas Kiani[*], Henry C. Frankis, Cameron M. Naraine, Dawson B. Bonneville, Andrew P. Knights and Jonathan D. B. Bradley

Department of Engineering Physics, McMaster University, 1280 Main Street West, Hamilton, ON L8S 4L7, Canada.

[*]e-mail: miarabbk@mcmaster.ca

**Abstract**
Silicon photonics technology enables compact, low-power and cost-effective optical microsystems on a chip by leveraging the materials and advanced fabrication methods developed over decades for integrated silicon electronics[1,2]. Silicon photonics fabrication foundries have evolved to provide services and process design kits containing the standard building blocks required for high-performance optical circuits, including passive components such as optical waveguides, filters and (de-)multiplexors and active optoelectronic components such as high-speed modulators, switches and photodetectors[3–5]. However, because silicon is a poor light emitting material, on-chip light sources are still a significant challenge for foundry offering in terms of the required provision for high-yield, widely deployed, integrated building blocks[6–8]. Integrated lasers are not offered in standard silicon photonics processes despite progress on various approaches using hybrid gain materials on silicon[9]. Current light-source integration methods are viewed as complex, requiring incompatible and/or expensive materials and processing steps. Here we report on an ultra-compact silicon photonic laser consisting of a thulium-silicon hybrid microdisk resonator. The microdisk design is straightforward and compatible with the fabrication steps and device dimensions available in all silicon photonics foundries, whereas the gain medium is added in a backend (final step), room temperature sputter deposition. This approach allows for low-cost and high-volume wafer-scale manufacturing and co-integration of light sources with silicon passive and active devices with no adjustment to standard process flows. The hybrid laser is pumped at standard telecom wavelengths around 1.6 μm and emits around 1.9 μm, which is within an emerging spectral region of significant interest for communications, nonlinear and quantum optics, and sensing on silicon[10–12].

**Introduction**
The development of integrated light sources is a persistent challenge in the field of silicon photonics. Silicon is an inefficient light-emitting material because it is an indirect bandgap semiconductor, thus electron-hole recombination must be mediated by phonon emission or absorption to conserve momentum. This makes radiative carrier recombination much less likely than non-radiative recombination. Further, light emission via carrier recombination is also limited to wavelengths < 1.1 μm due to silicon's 1.1 eV bandgap, which is below wavelengths compatible with low-loss silicon photonic microsystems. Various methods have been developed to build light sources onto silicon chips, including stimulated Raman emission in silicon[13], heterogeneously bonded or directly grown III-V semiconductors[14], germanium[15] and germanium-tin alloys[16], silicon-nitride based parametric oscillators[17], and silicon-organic hybrid devices[18]. While many promising results have been reported, each method suffers from one or more significant drawbacks, including high fabrication complexity, high cost, low yield, and/or inefficient performance under standard operating conditions. Therefore, in most applications, the light source is still located externally from the silicon chip.

Rare-earth-doped optical gain materials have recently attracted interest for silicon-based light sources because they can be deposited monolithically on silicon allowing wafer-scale fabrication, are low-cost, enable straightforward design, have high thermal stability and provide broadband optical gain and lasing in near-infrared bands of interest[19]. Early development of rare-earth light sources on silicon photonics platforms focused on introducing rare-earth dopants directly in silicon but were hampered by its low rare-earth solubility[20]. This led to hybrid designs based on fiber-coupled microcavities[21,22] or integration of rare-earth-doped materials with silicon-compatible materials such as silicon nitride[23,24]. Extending this approach to the development of on-chip lasers integrated with silicon has proven challenging because of a combination of relatively low rare-earth gain and high optical loss in a silicon waveguide, making it difficult to achieve roundtrip net optical gain and lasing in silicon-based cavities. Recently, rare-earth waveguide lasers have been demonstrated on a silicon-on-insulator (SOI) silicon photonics platform using a hybrid design based on a silicon nitride cavity and erbium-doped aluminum oxide coatings[25]. However, there are significant drawbacks of this approach, including the lack of integration directly in the silicon layer, the relatively large device size, and the requirement for high-power optical pumping. Such rare-earth laser designs require multiple low-loss silicon nitride waveguide layers in addition to the silicon waveguide layer, which increases the design complexity, is a non-standard feature in many silicon photonics foundries and limits access to the efficient active functionality possible within the silicon layer, such as modulation and detection. Further, the device footprint is orders of magnitude larger than other silicon photonic devices, with cavity lengths of several millimeters, limiting integration density in silicon photonic microsystems. The requirement for an off-chip optical pump would also limit their scalability and utility, unless smaller designs can enable one optical pump source to power many lasers on the chip. Lasing in a more compact and straightforward design directly in silicon would enable a low-cost, scalable approach to rare-earth lasers on silicon.

Here, we demonstrate a thulium-silicon hybrid-integrated microdisk laser. Compared to other silicon-based light sources the laser structure is straightforward, robust, scalable, low-cost and can be implemented using existing wafer-scale silicon photonics fabrication processes and a single room-temperature post-processing step. In contrast to previous rare-earth lasers, the hybrid microdisk design is exceedingly simple, the laser cavity and output are directly in the silicon layer, and the ultra-compact device size of 40-μm-diameter is on a scale compatible with standard passive and active silicon photonic devices. We show lasing with thulium as a rare-earth dopant, via efficient pumping at wavelengths around 1.6 μm, where silicon is highly transparent and off-chip or bonded on-chip light sources are readily available and could power many integrated devices. Single-mode emission is observed at extended near infrared wavelengths around 1.9 μm. Besides offering a low-cost integrated light source for silicon photonic microsystems, such lasers provide an incentive for expanding applications in an emerging 2-μm wavelength band[26], motivated by silicon's lower two-photon absorption and the recent development of efficient monolithic passive and active silicon devices in this range[27–30]. Optical gain and lasing using a hybrid silicon rare-earth structure opens the door to highly compact monolithic optical amplifiers as well as new types of effective solid state light sources (e.g., tunable lasers), using the active functionality in the silicon layer, on silicon photonics platforms.

**Results**
In Fig. 1, we display the microdisk laser structure. The hybrid laser consists of an integrated silicon microdisk and bus waveguide which are coated in a thulium-doped tellurium oxide (TeO$_2$:Tm$^{3+}$) layer. The silicon structure was fabricated in a silicon photonics foundry on a wafer-scale SOI platform with a 220-nm silicon layer thickness and consists of a 40-μm-diameter silicon microdisk next to a point-coupled silicon bus waveguide. After silicon waveguide etching, and deep etching for end-facet preparation and wafer dicing, the uncoated SOI chips were transferred from the foundry for post-processing. We then coated the entire structure in a 0.37-μm-thick thulium-doped tellurium oxide ($TeO_2:Tm^{3+}$) film using a room-temperature reactive co-sputtering post-processing step. Further details on the fabrication can be found in the Methods section. Tellurium oxide was selected as a host material for thulium because of its low loss, high rare-earth solubility, high refractive index (n ~2.1) for compact devices and enhanced mode overlap, good thermal and chemical stability, and low temperature deposition, allowing for straightforward post-processing on chips containing silicon photonics active layers including metallic interconnects[31]. Of various rare-earth elements, we selected thulium because it offers key advantages as a laser ion for silicon. Trivalent thulium ions are a quasi-three level laser system with thermally-populated broadened Stark-split ground and excited states, which can be pumped at telecom wavelengths around 1.6 μm and show broad emission from ~1.7–2.1 μm on the $^3F_4$ excited state to $^3H_6$ ground state energy transition. $Tm^{3+}$ ions exhibit relatively shifted absorption and emission spectra (resulting in low ground state absorption at longer laser wavelengths), minimal quenching effects at high ion concentrations and emission near the edge of silicon's low two-photon absorption window, allowing for straightforward population inversion in the laser cavity and efficient optical pumping and lasing. However, while high efficiency on-chip thulium lasers have been demonstrated in crystalline waveguides[32] and dielectric host materials on silicon substrates[33–35], thulium is relatively unexplored as a laser ion in an SOI platform. Importantly, here a standard silicon photonics foundry process was used to fabricate the disk platform and bus waveguide, which enables such hybrid lasers to be co-integrated with other passive and active silicon photonic devices on the same chip. Figure 1a shows a 3D drawing of the hybrid laser and energy level diagram for $Tm^{3+}$ with the energy bands and transitions most relevant for the laser operation. A cross section diagram and a focused-ion-beam (FIB)-milled cross section of the silicon laser structure are displayed in Fig. 1b and 1c, respectively. Figure 1d shows a top view scanning electron microscope (SEM) image of the hybrid laser.

We summarize the calculated optical properties of the hybrid $TeO_2:Tm^{3+}$-Si resonator structure in Fig. 2. The electric field profiles of the transverse-electric- (TE-) polarized fundamental modes for the microdisk resonator and waveguide and 1610-nm pump wavelength are displayed in Fig. 2a and Fig. 2b, respectively. Similarly, the TE-polarized fundamental modes for the microdisk and waveguide at the laser wavelength of 1906 nm are shown in Fig. 2c and Fig. 2d, respectively. To achieve gain in the disk resonator structure, a sufficient percentage of optical intensity must overlap with the $TeO_2:Tm^{3+}$ gain material. The theoretical properties of the pump and laser modes in the microdisk structure are displayed in Fig. 2e. For the microdisk resonator, approximately 15% and 19% of the optical power is confined in the $TeO_2:Tm^{3+}$ coating at the pump and laser wavelengths, respectively. Because of strong optical confinement in the silicon disk, low loss and a relatively high $Tm^{3+}$ concentration are required for sufficient amplification in the resonator and to achieve roundtrip net gain and lasing. The propagation loss in the microdisk includes contributions from the Si, $TeO_2$ and $SiO_2$ linear absorption, Si nonlinear (two-photon) absorption, $Tm^{3+}$ absorption, scattering loss at interfaces and bending radiation loss. Linear absorption is low for all materials used in the hybrid resonator. The two-photon absorption coefficient at 1.9 μm is about half that at 1.5 μm[36], thus nonlinear absorption and its influence on the roundtrip gain are expected to be minimal near the lasing threshold. Since $Tm^{3+}$ absorption at the laser wavelength is also low because of the small $Tm^{3+}$ absorption cross section around 1900 nm, in our design we expect to be limited by the bending radiation and scattering loss, the former of which can be designed to be low by selecting an appropriate bending radius and the latter of which is limited by fabrication steps such as etching and must be extracted from experiment. We calculated the theoretical radiation loss and equivalent $Q$ factor for the $TeO_2$:$Tm^{3+}$-coated silicon microdisk structure using a finite element bent eigenmode solver, as shown in Fig. 2f. The calculated radiation limited $Q$ factor at pump and laser wavelengths shows that radiation loss is negligible at the selected bend radius of 20 μm. These results show that such disks can be effectively designed with a 20-μm radius and below before introducing significant radiation losses, potentially allowing for the fabrication of more compact devices than reported here. When the bending loss is negligible the internal $Q$ factor of the microdisk becomes limited by absorption and scattering losses, which are independent of bend radius. The dashed lines in Fig. 2f indicate the internal $Q$ factors corresponding to different absorption/scattering-limited microdisk propagation losses. To achieve roundtrip gain and lasing in the cavity, the gain must exceed the cavity losses, including radiation, absorption/scattering loss and resonator-waveguide coupling loss. The inset in figure 2f shows the calculated upper limit of the Tm gain for different Tm concentrations, including the Tm concentration selected in this study. These results show that net roundtrip gain is achievable, accounting for the fact that the resonators are point-coupled and in the under-coupled regime and the coupling coefficient is low. Additional details regarding these calculations can be found in the Methods section.

We characterized the passive transmission properties of the microdisk resonator using a tunable laser and a fiber probe station (see Methods section for further details). As displayed in Fig. 3a, we observe narrow resonances associated with five different TE modes supported by the microdisk resonator. The extinction ratios are relatively low (≤ 1 dB), indicating the resonator is in the under-coupled regime. In this regime higher roundtrip gain can be achieved, increasing the probability of lasing, with the tradeoff that lower pump-power is coupled into the resonator resulting in reduced efficiency. By fitting the transmission responses of the under-coupled resonator using a Lorentzian function (inset of Fig. 3a), we obtain an internal quality factor, $Q_i$, of $5.6 \times 10^5$ at 1521 nm. This $Q_i$ corresponds to a propagation loss of 1.0 dB/cm in the microdisk cavity. We expect even lower background propagation loss at the laser wavelength around 1900 nm[37]. We also obtain an internal quality factor of $2.7 \times 10^5$ corresponding to 1.9 dB/cm propagation loss at the pump wavelength of 1610 nm, confirming the higher $Tm^{3+}$ ion absorption loss at this wavelength. These measurements are also shown on the plot in Fig. 2f and confirm that the $Q$ factor is limited by $Tm^{3+}$ absorption, background material loss and scattering loss inherent to the hybrid disk structure. In Fig. 3b, we measured the internal quality factors for various resonant modes and associated free spectral ranges (FSRs) (inset of Fig. 3b), in transmission experiments from 1520 nm to 1620 nm wavelength. The various microdisk modes show $Q$ factors with similar magnitude and wavelength dependence, indicating similar background loss and mode overlap with the silicon and $TeO_2$:$Tm^{3+}$ layers. The relatively large FSR of > 5 nm for our design increases the likelihood of single-mode as opposed to multi-mode lasing observed in previous integrated Tm microcavity lasers[33] and allows for engineering of the laser output wavelength by small adjustments in the resonator dimensions. The FSRs match well with those predicted via calculation, and allow for us to confirm pumping on the fundamental TE microdisk mode.

We describe the $TeO_2:Tm^{3+}$-Si hybrid microdisk laser results in Fig. 4. A schematic of the measurement setup used to characterize the lasers is displayed in Fig. 4a. We resonantly pumped the $TeO_2:Tm^{3+}$-coated Si microdisk resonators to investigate their lasing potential, with up to 32.4 mW power launched into the bus waveguide. Lasing was observed for a disk-bus gap of 0.6 μm and the highest slope efficiency and laser output power were observed for pumping at the fundamental TE mode around 1610 nm. The laser output is bidirectional and we observe similar output power at the pump input side of the chip. As shown in Fig. 4b, we observe a single-mode silicon microdisk laser output spectrum with up to 580 μW single-sided on-chip power and 1.16 mW double-sided output power. The inset shows the pump transmission and laser output spectrum measured at the chip output. A single laser line at 1906 nm is evident with a side-mode suppression of > 30 dB. We observe the laser output signal to be highly stable at room temperature without thermal control of the substrate or any adjustment of the pump wavelength required, even though the device is pumped on a narrow resonance. This is a result of silicon's relatively strong thermo-optic effect which provides a natural closed-loop system for our microdisk laser when pumped on the blue-side of resonance. Stable operation was observed for at least 9 hours under 29 mW of pumping, with no external thermal stabilization of the laser chip. As shown in Fig. 4c, we observed lasing at a threshold pump power of 16 mW launched into the bus waveguide and 2.5 mW coupled into the microdisk resonator. The single-sided laser slope efficiency versus launched and microdisk-coupled pump power is 4.2 % and 30 %, respectively, which yields a bidirectional slope efficiency with respect to absorbed pump power of up to 60%.

With this result we have demonstrated a simple compact monolithic laser on silicon. Compared to prior work on hybrid rare-earth silicon lasers[22], our entire structure is integrated on-chip with light emission in a silicon photonic waveguide and the fabrication methods are fully compatible with a silicon photonics foundry process. Furthermore, compared to recent rare-earth lasers on silicon photonics platforms based on silicon nitride cavities, the laser design is significantly less complex[25], has a 25 times smaller footprint in terms of area[33], and is directly integrated on the silicon layer. Integration of rare-earth gain materials with silicon can enable active optoelectronic or highly efficient thermo-optic laser control via doping the silicon or implementing metal heaters, respectively. Our compact design allows for straightforward and large-scale integration of such lasers within silicon photonic circuits, where potentially one pump could power multiple lasers (e.g., for wavelength division multiplexing applications or parallel sensors) by partially coupling pump power to each laser and efficiently tuning each laser onto resonance. By showing optical gain and lasing in a hybrid rare-earth silicon structure we also build on recent advances on monolithic rare-earth-doped amplifiers on silicon platforms[38,39] and provide incentive for developing new types of ultra-compact optical amplifiers, which are another missing element in commercial silicon photonics process design kits. In addition to thulium devices, these results also provide a guide for further work on other rare-earth-silicon hybrid lasers and amplifiers operating in a wide range of wavelengths.

**Conclusions**

We have demonstrated a single-mode hybrid thulium-silicon laser integrated on a chip. The hybrid laser is ultra-compact and fabricated using standard silicon photonic processing methods and cost-effective, low-temperature and wafer-scale post-processing steps enabling large-scale integration, volume production and implementation within advanced silicon photonic microsystems. The laser emits at 1906 nm with a threshold pump power of 16 mW and 2.5 mW with respect to the power coupled into the silicon bus waveguide microdisk, respectively. Considering bidirectional emission, we observe more than 1 mW total on-chip output power and a slope efficiency versus absorbed pump power of 60%. This new, stable, compact, inexpensive, efficient, room temperature silicon laser has implications for ultra-compact light sources for silicon-based photonic microsystems in the emerging 2-μm wavelength band. Integrated silicon lasers are in high demand for applications including data communications, quantum information systems, artificial intelligence, nonlinear optical systems, mid-infrared light generation, humidity, gas and bio-sensors, detection and ranging, spectroscopy, and advanced metrology.

**Acknowledgements**
We thank CMC Microsystems and the SiEPIC Program for facilitating the silicon photonics fabrication and the Centre for Emerging Device Technologies (CEDT) at McMaster University for support with the reactive sputtering system. We acknowledge financial support from the Natural Sciences and Engineering Research Council of Canada (grant numbers RGPIN-2017-06423, STPGP 494306 and RTI-2017-00474), the Canadian Foundation for Innovation (CFI project number 35548) and the Ontario Research Fund (ORF project numbers 35548 and RE-09-051).

**Author contributions**
K.M.K., H.C.F., C.M.N., D.B.B., A.P.K. and J.D.B.B. conceived of the idea of the project and contributed to the device design. H.C.F., C.M.N. and D.B.B. laid out the silicon chips. H.C.F. developed the low-loss tellurium oxide deposition recipe. K.M.K developed and optimized the thulium-doped tellurium oxide fabrication process, deposited the thulium-doped films on the silicon chips and performed the experimental characterization and analysis. K.M.K. and J.D.B.B. wrote the manuscript, with input from A.P.K. A.P.K. and J.D.B.B. supervised and coordinated the project. All authors commented on the manuscript.

**Additional information**
Supplementary information is available in the online version of the paper. Reprints and permissions information is available online at www.nature.com/reprints. Correspondence and requests for materials should be addressed to K.M.K.

**Competing financial interest**
The authors declare no competing financial interests.

**Methods**
**Laser Fabrication.** The laser chips were fabricated on an SOI platform using the Advanced Micro Foundry (AMF) silicon photonics fabrication process in Singapore. Silicon strip bus waveguides of 0.45 µm width and microdisks with radii of 20 µm and gaps varying from 0.2–1.6 µm between the outer walls of the silicon disk and bus waveguide, were patterned using deep ultraviolet 193-nm lithography into a 0.22-µm-thick silicon waveguide layer on a 2 µm-thick $SiO_2$ buried oxide (BOX) layer, without top $SiO_2$ cladding, allowing for post-process $TeO_2:Tm^{3+}$ thin film deposition. The silicon waveguides were tapered from 0.45- to 0.18-µm width at the edge of the chip and deep trenches were etched into the BOX and silicon handle wafer to allow for low-loss fiber-chip light coupling. The SOI wafer was diced along the deep trenches into chips and the chips were transferred from the foundry. We then deposited a 0.37-µm-thick $TeO_2:Tm^{3+}$ coating layer onto the passive silicon chips at McMaster University using a radio frequency (RF) reactive magnetron co-sputtering process. The process is similar to that reported for fabrication of silicon nitride hybrid amplifiers[40]. Three-inch metallic tellurium and thulium targets with 99.999 and 99.9% purity, respectively, were sputtered in an argon/oxygen atmosphere at ambient temperature. We set the Te and Tm RF sputtering powers to 120 W and 85 W, respectively, and the Ar and $O_2$ flow rates to 12 sccm and 9.4 sccm, respectively, at 20°C. The deposition rate for the $TeO_2:Tm^{3+}$ film was 13 nm/min and its refractive index was 2.04 at 638 nm and 1.99 at 1550 nm wavelengths measured by spectroscopic ellipsometry. TeO$_2$:Tm$^{3+}$ thin film propagation losses of ≤ 0.5 dB/cm at 1510 nm were determined using the prism coupling method and a separate film prepared on a thermally oxidized silicon substrate in the same deposition run. We measured a thulium ion dopant concentration of 4.0 × 10$^{20}$ cm$^{-3}$ using Rutherford backscattering spectrometry (RBS). The thulium ion dopant concentration of 4.0 × 10$^{20}$ cm$^{-3}$ was selected to be high enough to achieve greater gain than microdisk roundtrip losses, including propagation and disk-waveguide coupling losses. It is important to note that although the chips were unclad and passive-only in this study, the laser design allows for similar processing steps to be carried out on silicon photonic chips with active device layers, including metals and dopants, by etching a window into the top SiO$_2$ cladding (a standard processing step available within the silicon photonics foundry) and due to the low temperature TeO$_2$:Tm$^{3+}$ deposition.

**Laser Loss and Gain Calculations.** We calculated the electric field profiles and intensity overlap factors of the hybrid TeO$_2$:Tm$^{3+}$-silicon waveguide and microdisk modes and radiation losses of the microdisk using a finite-element method modesolver (RSoft FemSIM) and the cross-sectional structure shown in Fig. 1b. The microdisk cavity $Q$ factor was converted to an equivalent propagation loss using the assumptions and method outlined in reference 41. We calculated the maximum Tm$^{3+}$ gain achievable in the resonator, $\gamma_{\text{Tm}}$ (in dB/cm), using the following equation:

$$\gamma_{\text{Tm}} = 10 \log_{10} e \times \Gamma \times \sigma_{21} \times N_{\text{Tm}}$$

where $\Gamma$ and $\sigma_{21}$ are the intensity overlap factor and Tm$^{3+}$ ion emission cross section on the $^3F_4$–$^3H_6$ transition estimated for tellurite glass from reference 42, respectively, at the laser wavelength, $N_{\text{Tm}}$ is the Tm$^{3+}$ ion concentration and we assume full Tm$^{3+}$ population inversion to give an upper limit to the gain. We note that by using similar calculations for the hybrid bus waveguide structure, the total thulium-related pump absorption and laser signal enhancement in the waveguide were determined to be < 1 dB and have minimal impact on the pump threshold power and laser output power, respectively.

**Laser Measurements.** We characterized the passive and active properties of the silicon microdisk laser using the experimental setup shown in Fig. 4a. We coupled polarized pump light from a tunable 1520–1640 nm laser set at ~1610 nm and a high-power L-band erbium-ytterbium-co-doped fiber amplifier (L-band EYDFA) to the chip via a fiber polarization controller, a 1600/1900 nm fiber wavelength division multiplexor (WDM), and a 2-μm spot size at 1550-nm wavelength tapered optical fiber mounted on an *xyz* alignment stage. The microdisk laser output was also coupled from the chip using a tapered fiber mounted on an *xyz* stage, filtered from the pump light with a 1600/1900 nm WDM, and coupled to an optical spectrum analyzer (OSA) to observe the output spectrum. The transmitted pump light was also measured using an optical power meter. During measurements the polarization paddles and *xyz* stages were adjusted to select TE polarization and maximize the off-resonance transmitted pump/signal intensity. The fiber-chip coupling loss at 1610 nm was determined to be 7.5 dB, influenced by mode mismatch, Fresnel reflections and scattering due to the conformal TeO$_2$:Tm$^{3+}$ coating on the facet. We determined the launched pump power by measuring the incident power from the input fiber using an integrating sphere photodiode power monitor and accounting for fiber-chip coupling loss.

Transmission measurements were carried out on the same setup without the L-band EYDFA and the OSA replaced with a photodetector to determine the background microdisk propagation loss. We measured the transmission in the wavelength range of 1520 to 1540 nm where we observe negligible thulium absorption loss and the fitted intrinsic $Q$ factor can be assumed to represent the passive propagation loss of the structure.

Figure 1

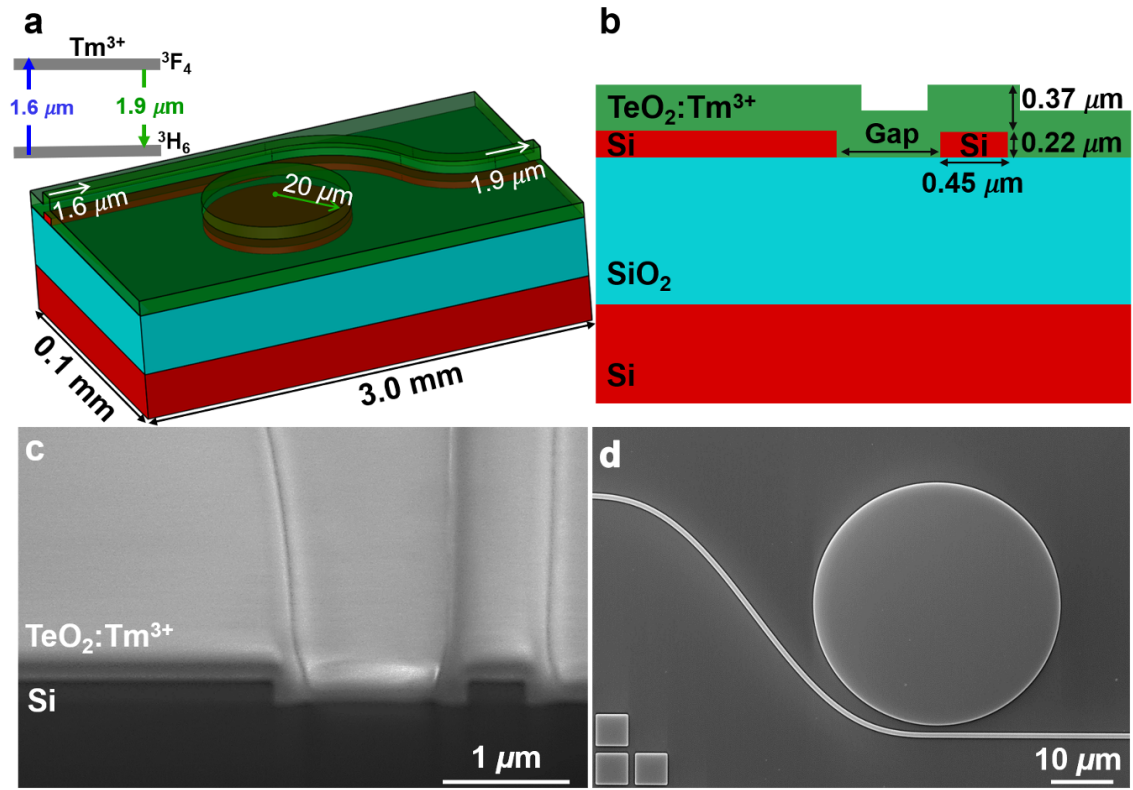

**Figure 1 | Silicon hybrid microdisk laser images. a,** 3D drawing of the TeO$_2$:Tm$^{3+}$-coated silicon microdisk laser. Inset in a: Two-level Tm$^{3+}$ energy diagram showing 1.6 μm photon absorption and excitation into upper level and de-excitation into the lower level and 1.9 μm photon emission. **b,** Cross-section profile of the TeO$_2$:Tm$^{3+}$-coated Si microdisk laser showing the microdisk structure and the bus waveguide dimensions. **c,** Focused ion beam milled scanning electron microscope (SEM) cross-section image of the coupling region between the TeO$_2$:Tm$^{3+}$-coated Si microdisk and bus waveguide. **d,** Top-view SEM image of a TeO$_2$:Tm$^{3+}$-coated Si microdisk laser.

Figure 2

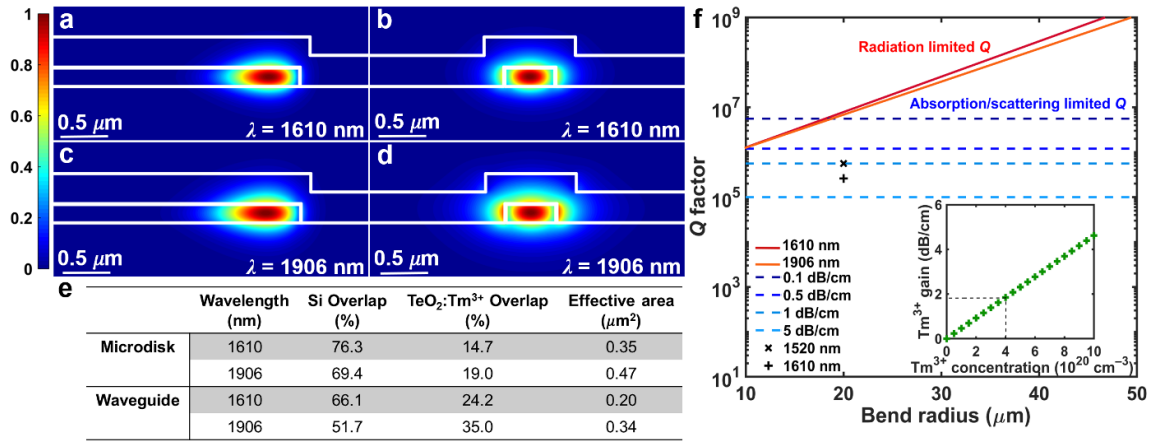

**Figure 2 | Calculated pump and laser modes and microdisk resonator loss and gain.** Calculated electric field profile of the fundamental transverse-electric- (TE-) polarized mode for the laser wavelength using a finite-element method mode solver for **a**, the $TeO_2$:$Tm^{3+}$-coated silicon microdisk and **b**, the $TeO_2$:$Tm^{3+}$-coated silicon strip waveguide at the 1610 nm pump wavelength and **c**, microdisk and **d**, waveguide at the 1906 nm lasing wavelength. **e**, calculated fractional optical intensity overlaps and effective mode areas for the fundamental TE microdisk mode at the 1610 nm pump and 1906 nm lasing wavelengths. **f**, Calculated internal $Q$ factor of the hybrid silicon microdisk laser at the pump and lasing wavelengths. The measured $Q$ factors at 1521 and 1610 nm wavelength and 20 µm bend radius are also indicated with lines shown for several resonator losses, showing that the internal $Q$ is limited by the absorption and scattering related propagation loss of the microdisk. The inset shows the calculated upper limit $Tm^{3+}$ gain versus thulium concentration in the hybrid resonator.

**Figure 3**

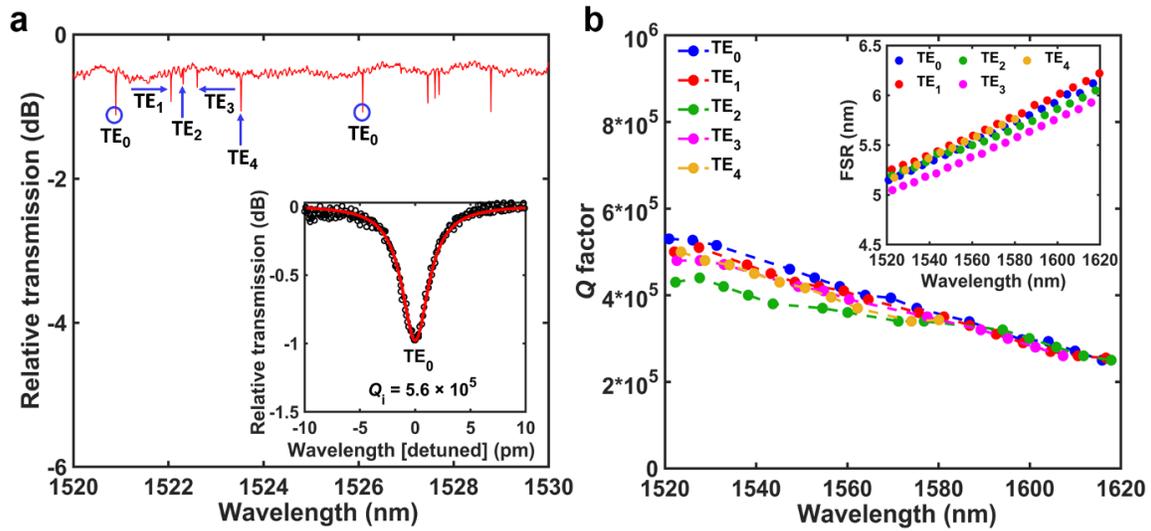

**Figure 3 | Transmission and loss measurements. a,** Measured TE transmission spectrum for a TeO$_2$:Tm$^{3+}$-coated silicon microdisk with a microdisk-waveguide gap of 0.6 µm. Inset in a: a close-up view of the under coupled resonance at 1521 nm wavelength with extinction ratio of 0.95 dB and a fitted Lorentzian function yielding an intrinsic quality factor, $Q_i$, of 5.6 x 10$^5$ corresponding to 1 dB/cm background optical propagation loss. **b,** Microdisk intrinsic quality factor versus wavelength confirming the onset of thulium absorption at longer wavelengths. Inset in b: free spectral ranges (FSRs) of the microdisk modes versus wavelength.

**Figure 4**

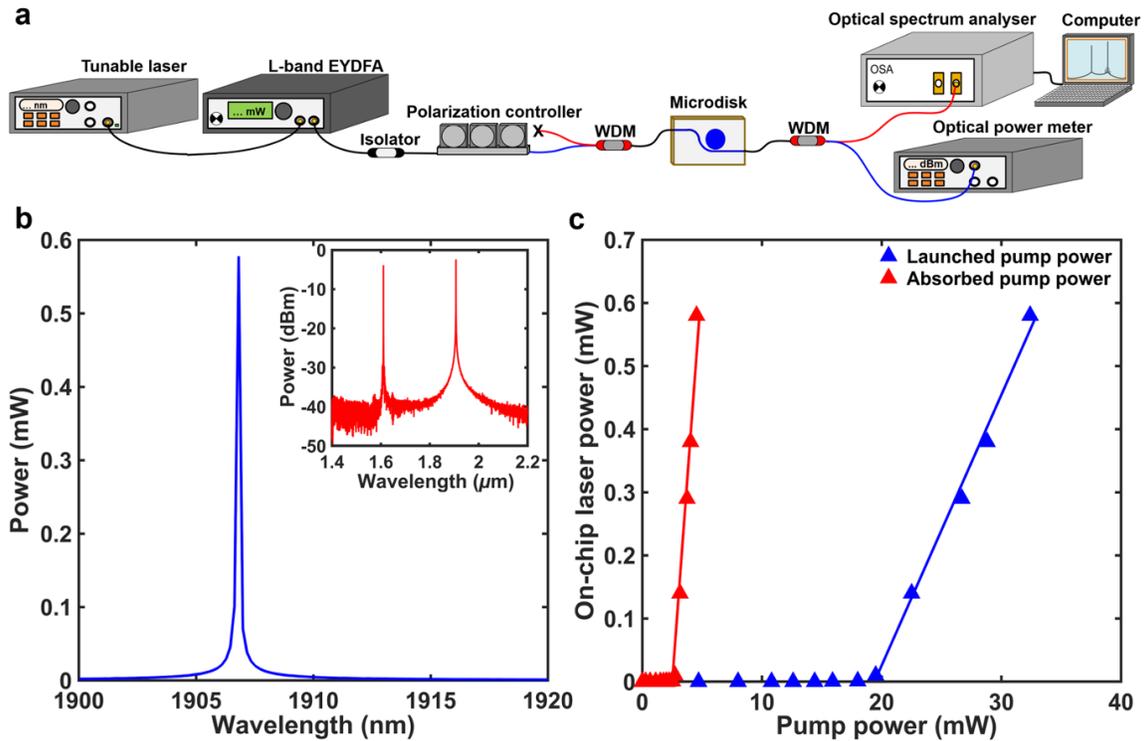

**Figure 4 | Hybrid thulium-silicon microdisk laser measurements. a,** Experimental setup used for measuring the on-chip hybrid $TeO_2$:$Tm^{3+}$-coated silicon microdisk lasers. **b,** Single mode laser emission spectrum at 1906 nm of a $TeO_2$:$Tm^{3+}$-coated Si microdisk resonator under 1610-nm pumping at a microdisk-waveguide gap of 0.6 μm obtained with 32.4 mW on-chip pump power. Inset in **b**: laser emission spectrum with a side-mode suppression of > 30 dB. **c,** Laser output curve showing a single-sided on-chip output power of up to 580 μW, slope efficiencies of 4.2% and 30% versus on-chip launched and absorbed 1610 nm pump power, and threshold pump powers of 16 mW in the bus waveguide and 2.5 mW in the microdisk resonator.